\documentclass[letterpaper,11pt]{article}

\usepackage{jheppubMod}
\usepackage{adjustbox}
\usepackage{academicons}      
\usepackage{array,multirow}
\usepackage{bm}  %
\usepackage{booktabs}
\usepackage{float}
\usepackage[T1]{fontenc} 
\usepackage[symbol]{footmisc}
\usepackage{braket,bm}
\usepackage{mathrsfs}
\usepackage{mathtools}
\usepackage{slashed,soul}
\usepackage{tikz}
\usetikzlibrary{arrows,decorations.pathmorphing,backgrounds,positioning,fit,petri,automata,shadows,calendar,mindmap,
decorations.markings,calc}
\usepackage{xspace}
\usepackage{slashed}
\usepackage{array,multirow}
\usepackage{graphicx}
\usepackage{graphics}
\usepackage{epsfig}
\usepackage{amsfonts}
\usepackage{amsmath}
\usepackage{amssymb}
\usepackage{dsfont}
\usepackage{color}
\usepackage{pdflscape}
\usepackage{pifont}
\usepackage{pgfplots}
\usepackage{pgfplotstable}
\usepackage{bbold}
\usepackage[normalem]{ulem}
\usepackage{xcolor,cancel,youngtab}

\usepackage{hyperref}
\hypersetup{
    colorlinks=true,       
    linkcolor=blue,          
    citecolor=blue,        
    filecolor=blue,      
    urlcolor=blue           
}

\definecolor{labelkey}{rgb}{0,0.5,0.0}

\newcolumntype{P}[1]{>{\centering\arraybackslash}p{#1}}

\newcommand{\beq}{\begin{equation}}
\newcommand{\eeq}{\end{equation}}
\newcommand{\be}{\begin{equation}}
\newcommand{\ee}{\end{equation}}
\newcommand{\bea}{\begin{eqnarray}}
\newcommand{\eea}{\end{eqnarray}}
\newcommand{\ben}{\begin{eqnarray*}}
\newcommand{\een}{\end{eqnarray*}}

\newcommand{\bma}{\begin{pmatrix}}
\newcommand{\ema}{\end{pmatrix}}

\def\lixo#1{}

\def\slashchar#1{\setbox0=\hbox{$#1$}           
  \dimen0=\wd0                                    
  \setbox1=\hbox{/} \dimen1=\wd1                  
  \ifdim\dimen0>\dimen1                           
    \rlap{\hbox to \dimen0{\hfil/\hfil}}            
    #1                                             
  \else                                          
    \rlap{\hbox to \dimen1{\hfil$#1$\hfil}}        
    /                                           
 \fi}                                           %

\newcommand{\dslash}[1]{#1 \llap{/\kern-0.5pt}}
\newcommand{\Dslash}[1]{#1 \llap{/\kern+1.5pt}}
\newcommand{\DDslash}[1]{#1 \llap{/\kern+2.3pt}}
\newcommand{\dslashh}[1]{#1 \llap{/\kern+1pt}}

\definecolor{cadmiumgreen}{rgb}{0.0, 0.42, 0.24}
\definecolor{darkpastelgreen}{rgb}{0.01, 0.75, 0.24}
\definecolor{darkspringgreen}{rgb}{0.09, 0.45, 0.27}
\definecolor{forestgreen(web)}{rgb}{0.13, 0.55, 0.13}
\definecolor{forestgreen(traditional)}{rgb}{0.0, 0.27, 0.13}
\definecolor{cobalt}{rgb}{0.0, 0.28, 0.67}
\definecolor{darkblue}{rgb}{0.0, 0.0, 0.75}
\definecolor{darkred}{rgb}{0.55, 0.0, 0.0}
\definecolor{palatinatepurple}{rgb}{0.41, 0.16, 0.38}
\definecolor{burntorange}{rgb}{0.8, 0.33, 0.0}

\def\12{\frac{1}{2}}
\def\0nbb{$0\nu\beta\beta$}

\newcommand{\DLR}{\smash{\overset{\text{\small$\leftrightarrow$}}{\smash{D}\vphantom{+}}}}

\title{Beta-decay implications for the $W$-boson mass anomaly  }

\author[a]{Vincenzo Cirigliano,}
\author[a]{Wouter Dekens,}
\author[b,c]{Jordy de Vries,}
\author[d]{Emanuele Mereghetti,}
\author[e]{Tom Tong}

\affiliation[a]{Institute for Nuclear Theory, University of Washington, Seattle WA 91195-1550, USA}
\affiliation[b]{Institute for Theoretical Physics Amsterdam and Delta Institute for Theoretical Physics,\\ University of Amsterdam, Science Park 904, 1098 XH Amsterdam, The Netherlands}
\affiliation[c]{Nikhef, Theory Group, Science Park 105, 1098 XG, Amsterdam, The Netherlands}
\affiliation[d]{Theoretical Division, Los Alamos National Laboratory, Los Alamos, NM 87545, USA}
\affiliation[e]{Center for Particle Physics Siegen, University of Siegen, 57068 Siegen, Germany}

\abstract{
We point out the necessity to consider $\beta$-decay observables in resolutions of the $W$-boson anomaly in the Standard Model Effective Field Theory that go beyond pure oblique corrections. 
We demonstrate that present global analyses that explain the $W$-boson mass anomaly predict 
a large, percent-level, violation of first-row CKM unitarity. We investigate what solutions to the $W$-boson mass anomaly survive after including $\beta$-decay constraints.
\looseness-1
}

\keywords{}

\begin{document}
\maketitle
\setcounter{page}{2}
\flushbottom

\section{Introduction}

The recent announcement by the CDF collaboration of a new measurement of the W-boson mass $m_W=80433.5\pm 9.4$ MeV~\cite{CDF:2022hxs} is very exciting. Taken at face value, it implies significant tension with the Standard Model (SM) prediction~\cite{Haller:2018nnx} as well as with earlier, less precise, determinations~\cite{ATLAS:2017rzl,LHCb:2021bjt}. While only one week old, the announcement sparked a lot of interest to understand the implication of this measurement, 
if correct, on potential beyond-the-SM (BSM) physics. In particular, several groups studied the $W$-boson mass anomaly (as we will call this from now on) in terms of the Standard Model Effective Field Theory (SMEFT) under various assumptions to limit the number of independent operators and associated Wilson coefficients.

Most effort has been focused on explanations through oblique parameters \cite{Lu:2022bgw,Strumia:2022qkt,Paul:2022dds,Asadi:2022xiy} probing universal theories \cite{Kennedy:1988sn,PhysRevLett.65.964,Barbieri:2004qk}. Refs.~\cite{deBlas:2022hdk,Fan:2022yly,Bagnaschi:2022whn,Gu:2022htv,Endo:2022kiw,Balkin:2022glu} went beyond this approach by using a more general set of SMEFT operators. For example, Ref.~\cite{deBlas:2022hdk} fitted EWPO under the assumption of flavor universality, finding that the $W$-boson mass anomaly requires non-zero values of various dimension-six SMEFT Wilson coefficients. An important ingredient in these fits is the decay of the muon that enters in the determination of the Fermi constant. However, none of the analyses included the hadronic counterpart in the form of $\beta$-decay processes of the neutron and atomic nuclei~\cite{Falkowski:2020pma}, and semileptonic 
meson decays~\cite{Gonzalez-Alonso:2016etj}.
In this work, we argue that not including constraints from these observables in global fits generally leads to too large deviations in first-row CKM unitarity, much larger than the mild tension shown in state-of-the-art determinations. We investigate what solutions to the $W$-boson mass anomaly survive after including $\beta$-decay constraints.

\section{Mass of the W boson in SMEFT}

We adopt the parameterization of SMEFT at dimension-six in the \textit{Warsaw} basis~\cite{Weinberg:1979sa, Buchmuller:1985jz,Grzadkowski:2010es}.
\begin{equation}
	\mathcal{L}_\text{SMEFT}^{\text{dim-6}} =
	\mathcal{L}_\text{SM} +
	\sum_{i} C_i\mathcal{O}^{\text{dim-6}}_i \, ,
	\label{eq:LSMEFT}
\end{equation}
where $C_i$ are Wilson coefficients of mass dimension $-2$.

Calculated at linear order in SMEFT, the shift to $W$ mass from the SM prediction due to dimension-six operators is given by \cite{Berthier:2015oma,Bjorn:2016zlr}
\begin{equation}
	\frac{\delta m_W^2}{m_W^2} = 
	v^2 \ \frac{s_w c_w}{s_w^2 - c_w^2} 
	\left[ 2 \, C_{HWB} + \frac{c_w}{2 s_w} \, C_{HD} + 
	\frac{s_w}{c_w} \left( 2 \, C_{Hl}^{(3)} - C_{ll} \right) \right] \,,
\label{eq:mW}
\end{equation}
where $v\simeq 246$ GeV is the vacuum expectation value of the Higgs field, $s_w = \sin{\theta_w}$ and $c_w = \cos{\theta_w}$.
The Weinberg angle $\theta_w$ is fixed by the electroweak input parameters $\{ G_F, m_Z, \alpha_{EW} \}$ \cite{Brivio:2021yjb}. Here we define $\delta m_W^2 = m_W^2(\mathrm{SMEFT}) -m_W^2(\mathrm{SM})$. 
The mass of the W boson receives corrections from four Wilson coefficients, namely $C_{HWB}$, $C_{HD}$, $C_{Hl}^{(3)}$, and $C_{ll}$. For the corresponding operators, see Tab.~\ref{tab:SMEFT_Ops}.

$C_{HWB}$ and $C_{HD}$ are related to the oblique parameters $S$ and $T$ \cite{PhysRevLett.65.964}. They have been thoroughly studied for constraining 'universal' theories \cite{Barbieri:2004qk,Wells:2015uba} with electroweak precision observables as well as in light of the $W$-boson mass anomaly~\cite{Lu:2022bgw,Strumia:2022qkt,Paul:2022dds,Asadi:2022xiy}. The linear combination of Wilson coefficients shown in Eq.~\eqref{eq:mW}, $\left( 2 \, C_{Hl}^{(3)} - C_{ll} \right)$, is related to the shift to Fermi constant in SMEFT.

\begin{table*}[t]
	\centering
	\def\arraystretch{1.5}
	\begin{tabular}{|c|c|}
		\hline
		${\cal O}_{HWB}$ & $H^\dagger \tau^I H\, W^I_{\mu\nu} B^{\mu\nu}$ \\
		\hline
		${\cal O}_{HD}$ & $\big| H^\dagger D_\mu H \big|^2$ \\
		\hline
		${\cal O}_{Hl}^{(3)}$ & $\left(H^\dagger i \DLR^I_{\mu} H \right) \left(\bar l_p \tau^I \gamma^\mu l_r \right)$ \\
		\hline
		${\cal O}_{Hq}^{(3)}$ & $\left(H^\dagger i \DLR^I_{\mu} H \right) \left(\bar q_p \tau^I \gamma^\mu q_r \right)$ \\
		\hline
		${\cal O}_{ll}$ & $\left(\bar l_p \gamma_\mu l_r \right) \left(\bar l_s \gamma^\mu l_t \right)$ \\
		\hline
		${\cal O}_{lq}^{(3)}$ & $\left(\bar l_p \tau^I \gamma_\mu l_r \right) \left(\bar q_s \tau^I \gamma^\mu q_t \right)$ \\
		\hline
	\end{tabular}
	\caption{List of the most relevant SMEFT dimension-six operators that are involved in this analysis.}
	\label{tab:SMEFT_Ops}
\end{table*}

\section{EWPO fits and CKM unitarity}

Under the assumption of flavor universality, 10 operators affect the EWPO at tree level, but only 8 linear combinations can be determined by data \cite{deBlas:2022hdk}. Following Ref.~\cite{deBlas:2022hdk}, these linear combination are written with $\hat C_i$ notation and given by $\hat C_{Hf}^{(1)} = C_{Hf}^{(1)} - (Y_f / 2) C_{HD}$, where $f$ runs over left-handed lepton and quark doublets and right-handed quark and lepton singlets, and $\hat C_{Hf}^{(3)} = C_{Hf}^{(3)} + (c_w / s_w) C_{HWB} + (c_w^2 / 4 s_w^2) C_{HD}$ where $f$ denotes left-handed lepton and quark doublets, and $\hat C_{ll} = (C_{ll})_{1221}$. Here $Y_f$ is the hypercharge of the fermion $f$. 

Ref.~\cite{deBlas:2022hdk} reported the results of their fits including the correlation matrix from which we can reconstruct the $\chi^2$. For concreteness we use their `standard average' results but our point would hold for the `conservative average' as well. 
To investigate the consequences of CKM unitarity on the fit, we will assume the flavor structures of the operators follow Minimal Flavor Violation (MFV) \cite{Chivukula:1987py, DAmbrosio:2002vsn}. That is, we assume the operators are invariant under a $U(3)_q\times U(3)_u\times U(3)_d\times U(3)_l\times U(3)_e$ flavor symmetry.
In addition, we slightly change the operator basis and trade the Wilson coefficient $\hat C_{ll}$  for the linear combination
\begin{equation}
 C_{\Delta}= 2 \left[ C_{Hq}^{(3)} - C_{Hl}^{(3)} + \hat C_{ll} \right]\,.
\end{equation}
We then refit the Wilson coefficients to the EWPO and obtain the results in the second column of Table~\ref{tab:SMEFT1}. In particular, we obtain 
\bea\label{deBlas}
C_{\Delta} = -\left(0.19\pm 0.09\right)\, {\rm TeV}^{-2}\,.
\eea

This combination of Wilson coefficients contributes to the violation of unitarity in the first row of the CKM matrix tracked by $\Delta_{\mathrm{CKM}} \equiv |V_{ud}|^2+|V_{us}|^2-1$, 
where we neglected the tiny $ |V_{ub}|^2$ corrections. Within the MFV assumption, we can write \cite{Cirigliano:2009wk} 
\begin{equation}
\Delta_{\mathrm{CKM}} =v^2 \left[ C_{\Delta} - 2\,C_{lq}^{(3)} \right] \,.
\end{equation}
The $C_{lq}^{(3)}$ operator that appears here does not affect EWPO and does not play a role in the fit of Ref.~\cite{deBlas:2022hdk}. If one assumes this coefficient to be zero, Eq.~\eqref{deBlas} causes a shift 
\begin{equation}\label{CKMfit}
\Delta^{\mathrm{EWfit}}_{\mathrm{CKM}} =-(0.012\pm0.005)\,,
\end{equation}
implying large, percent-level, deviations from CKM unitarity. 

Based on up-to-date theoretical predictions for $0^+ \rightarrow 0^+$ transitions and Kaon decays \cite{Cirigliano:2008wn, FlaviaNetWorkingGrouponKaonDecays:2010lot,Seng:2018yzq,Czarnecki:2019mwq,Hardy:2020qwl,Aoki:2021kgd,Seng:2021nar}, the PDG average indicates that unitarity is indeed violated by a bit more than two standard deviations  
\cite{Zyla:2020zbs}
\bea\label{PDG}
\Delta_{\mathrm{CKM}}  = -0.0015(7)\,,
\eea
but in much smaller amounts than predicted by Eq.~\eqref{CKMfit}. 
 This exercise shows that global fits to EWPO and the W mass anomaly that include BSM physics beyond 
the oblique parameters S and T, such as the one of Ref.~\cite{deBlas:2022hdk}, 
are  severely disfavored by $\beta$-decay data. While we did not repeat the fits of Refs.~\cite{Bagnaschi:2022whn, Balkin:2022glu}, the central values of their Wilson coefficients also indicate a negative percent-level shift to $\Delta_{\mathrm{CKM}}$, consistent with Eq.~\eqref{CKMfit}.

\begin{figure}[!t]
\begin{center}
\includegraphics[trim={1.2cm 0 0 0},clip, width=0.8\textwidth]{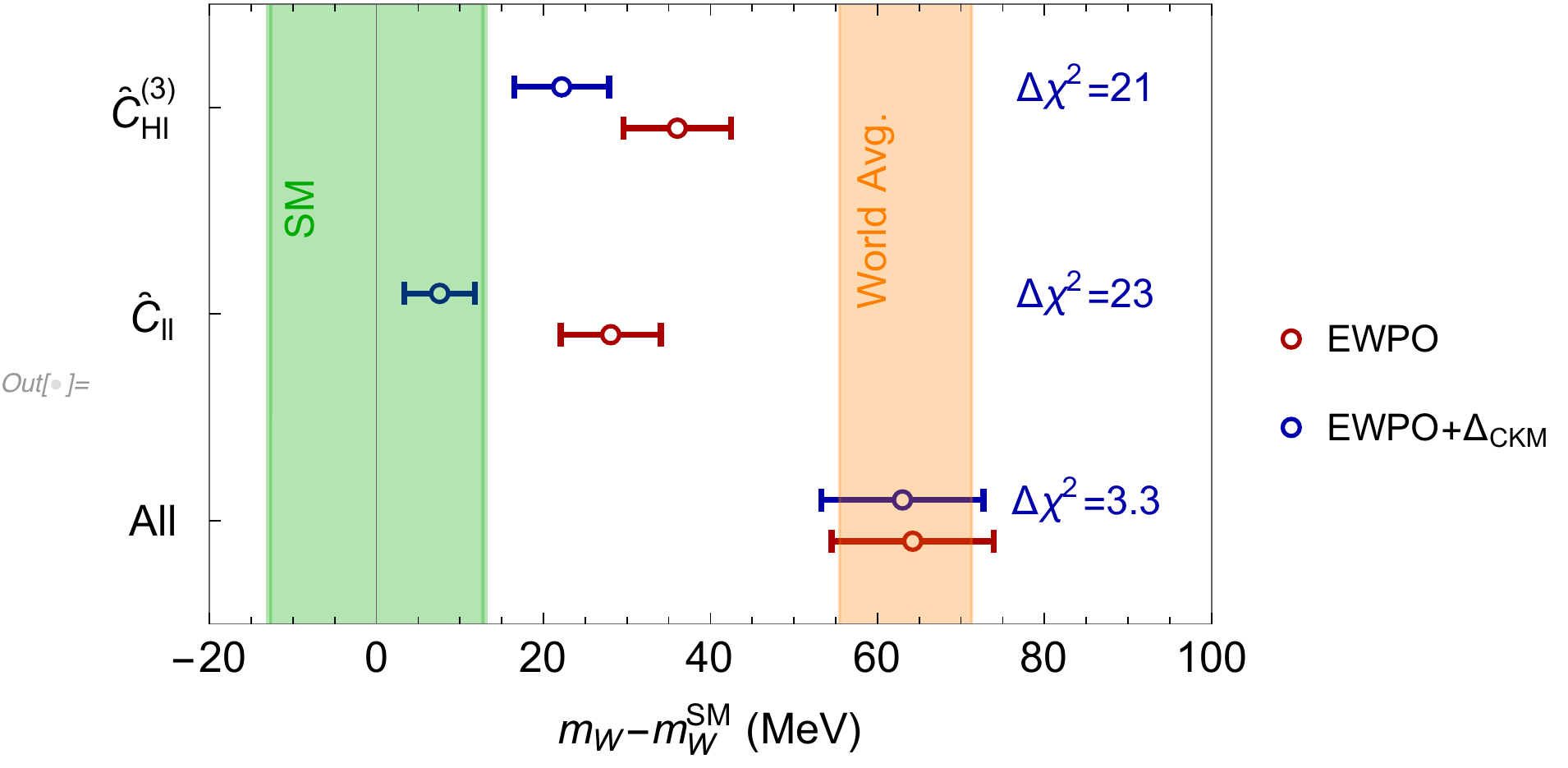}
\end{center}
\vspace{-0.5cm}
\caption{The resulting values of $\Delta m_W=m_W-m_W^{\rm SM}$ when turning on $\hat C_{Hl}^{(3)}$, $ \hat C_{ll}$, and all Wilson coefficients that are probed by EWPO. The red bars indicated the predicted $\Delta m_W$ from the EWPO fit, while the blue bars show the resulting $\Delta m_W$ after inclusion of $\Delta_{\mathrm{CKM}}$. The shown values of $\Delta \chi^2$, denote the differences in the minimum $\chi^2$ between the blue and red points. The SM prediction and world average, taken from Ref.\ \cite{deBlas:2022hdk}, are depicted by the green and orange bands, respectively. 
}\label{fig:mw}
\end{figure}

Indeed, combining the EWPO with $\Delta_{\rm CKM}$, we find that the minimum $\chi^2$ increases by $3.3$ and Wilson coefficients are shifted, as shown in Tab.~\ref{tab:SMEFT1}. 
Again this shows that the Cabibbo universality test has a significant impact and should 
be included in EWPO analyses of the $W$-boson mass anomaly. These statements are illustrated in Fig.~\ref{fig:mw}, 
which shows the values of 
$\Delta m_W=m_W-m_W^{\rm SM}$
obtained by fitting EWPO alone or EWPO {\it and} $\Delta_{\mathrm{CKM}} $
for two single-operator scenarios and the global analysis 
involving all operators.

\begin{table*}[t]
    \centering
    \begin{tabular}{c|c|c}
 \hline
 & Result & Result with CKM \\ 
 \hline
$\hat{C}_{\varphi l}^{(1)}$ & $ -0.007 \pm 0.011 $ &$ -0.013 \pm 0.009 $     \\ 
$\hat{C}_{\varphi l}^{(3)}$ & $ -0.042 \pm 0.015 $ & $-0.034\pm 0.014  $   \\ 
$\hat{C}_{\varphi e}$       & $ -0.017 \pm 0.009 $  &  $-0.021 \pm 0.009 $   \\
$\hat{C}_{\varphi q}^{(1)}$ & $ -0.0181\pm 0.044 $ &$  -0.048\pm 0.04   $   \\
$\hat{C}_{\varphi q}^{(3)}$ & $ -0.114 \pm 0.043 $  &$ -0.041\pm 0.015   $   \\
$\hat{C}_{\varphi u}$       & $ \phantom{+}0.086 \pm 0.154 $       & $-0.12\pm0.11 $    \\
$\hat{C}_{\varphi d}$       & $ -0.626 \pm 0.248 $   &  $-0.38  \pm 0.22 $   \\
$C_{\Delta}$          & $  -0.19 \pm 0.09 $     &  $ -0.027 \pm 0.011 $ \\ \hline
    \end{tabular}
    \caption{Results from the dimension-six SMEFT fit of Ref.~\cite{deBlas:2022hdk}, before and after the inclusion of $\Delta_{\mathrm{CKM}}$. All Wilson coefficients are given in units of TeV$^{-2}$. }
    \label{tab:SMEFT1}
\end{table*}

\begin{figure}[!t]
\begin{center}
\includegraphics[width=0.6\textwidth]{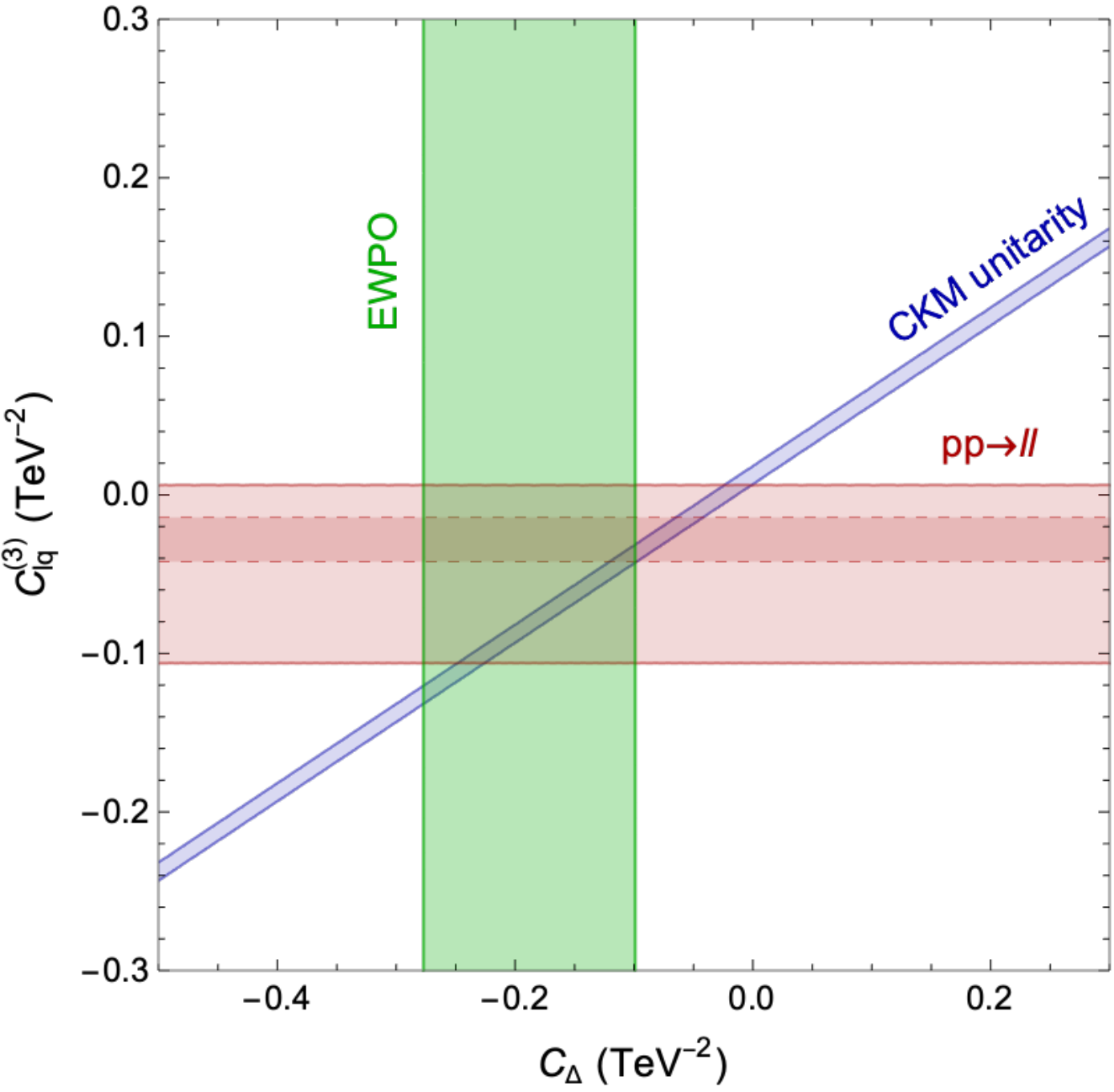}
\end{center}
\vspace{-0.5cm}
\caption{The $1\sigma$ constraints from EWPO in green, a global (single-coupling) analysis of LHC measurements in (dashed) red, and low-energy beta decays in blue. 
}\label{fig:plot}
\end{figure}

Another way to proceed is to effectively decouple the CKM unitarity constraint from EWPO by 
letting $C_{lq}^{(3)}\neq0$, which is 
consistent with the MFV approach. The $\Delta_{\mathrm{CKM}}$ observable is then accounted for by a nonzero value 
\begin{equation}
C_{lq}^{(3)} = -(0.082 \pm 0.045)\, {\rm TeV}^{-2}\,,
\end{equation}
while the values of the other Wilson coefficients return to their original value given in the second column of Table~\ref{tab:SMEFT1}. However, care must be taken that such values of $C_{lq}^{(3)}$ are not excluded by LHC constraints \cite{Bhattacharya:2011qm,Cirigliano:2012ab,Alioli:2018ljm,Panico:2021vav,Kim:2022amu}. 
In particular, Ref.\ \cite{Boughezal:2021tih} analysed
8 TeV $pp\to ll$ data  from \cite{ATLAS:2016gic} in the SMEFT at dimension-8.
Limiting the analysis to MFV dimension-six operators, we find 
\bea
C_{lq}^{(3)}&=& -(0.028\pm 0.028)\, {\rm TeV}^{-2}\quad ({\rm Single\, coupling},\quad 95\%\, {\rm C.L.})\,,\nonumber\\
C_{lq}^{(3)}&=& -(0.05\pm 0.1)\, {\rm TeV}^{-2}\quad\quad \,\,\,({\rm Global\, fit},\quad 95\%\, {\rm C.L.})\,,\label{eq:lhc}
\eea
when in the first line only $C_{lq}^{(3)}$ is turned on,
while in second line seven operators were turned on: 
$C_{lq}^{(1)}$, $C_{lq}^{(3)}$,
$C_{qe}$, $C_{lu}$, $C_{ld}$, $C_{eu}$, and $C_{ed}$.

The resulting constraints from EWPO, $\Delta_{\mathrm{CKM}}$, and the LHC are shown in Fig.\ \ref{fig:plot}. As mentioned above, a simultaneous explanation of $m_W$ and $\Delta_{\mathrm{CKM}}$ requires a nonzero value of $C_{lq}^{(3)}$, which implies effects in collider processes. The single-coupling bound from $pp\to ll$ in Eq.\ \eqref{eq:lhc} is already close to excluding the overlap of the EWPO and $\Delta_{\mathrm{CKM}}$ regions, while a global fit allows for somewhat more room. Nevertheless, should the current discrepancy in the EWPO fit hold, 
the preference for a nonzero $C_{lq}^{(3)}$ could be tested by
existing 13 TeV $pp \rightarrow l l$ \cite{ATLAS:2020yat}
and $pp \rightarrow l \nu$ data \cite{ATLAS:2019lsy}, and, in the future, at the HL-LHC.

\section{Conclusion}

In this note we have pointed out that global analyses of 
EWPO (beyond  oblique parameters) in the general SMEFT framework, while explaining the $W$-boson mass anomaly tend to predict 
a large, \% level, violation of  Cabibbo universality, parameterized  by  $\Delta_{\rm CKM}$.  
This result is not consistent with precision beta decay and meson decay phenomenology and 
calls for the inclusion of first-row 
CKM unitarity test in the set of EWPO, 
which is not commonly done. 
The inclusion of $\Delta_{\rm CKM}$ also requires 
adding $O_{lq}^{(3)}$ to the set of SMEFT operators 
usually adopted in EWPO analyses. 
We have illustrated this and shown that in this case 
Cabibbo universality can be recovered at the $0.1$\% level 
while still explaining the W mass anomaly. 
This extended scenario is currently 
consistent with constraints  on $C_{lq}^{(3)}$ from $pp \to ll$ at the LHC and will be tested / challenged  by future LHC data.

In this work we have only considered  the case of $U(3)^5$ flavor invariance  for the SMEFT  operators.  
Relaxing this hypothesis has several implications: 
first, one should consider flavor non-universal Wilson coefficients for the usual set of operators in EWPO fits; second, when including semileptonic low-energy processes, one should 
extend the operator set to include operators with more general Lorentz structures than $O_{lq}^{(3)}$, such as right-handed currents \cite{Alioli:2017ces}, that  provide additional ways to decouple the 
$\Delta_{\rm CKM}$ constraint from the W mass.  
The inclusion of flavor non-universal Wilson coefficients~\cite{Balkin:2022glu} in the EWPO fit 
has been shown to provide a good solution to the W mass anomaly. 
In this context, addressing the role of CKM unitarity constraints and lepton-flavor universality tests in meson decays is an interesting direction of future study.

\section*{Acknowledgements} 

VC and WD acknowledge support by the U.S. DOE under Grant No. DE-FG02-00ER41132.
JdV acknowledges support from the Dutch Research Council (NWO) in the form of a VIDI grant. 
EM is supported  by the US Department of Energy through  
the Office of Nuclear Physics  and  the  
LDRD program at Los Alamos National Laboratory. Los Alamos National Laboratory is operated by Triad National Security, LLC, for the National Nuclear Security Administration of U.S.\ Department of Energy (Contract No. 89233218CNA000001).
The work of TT is supported by the University of Siegen under the Young Investigator Research Group (Nachwuchsforscherinnengruppe) grant.

\bibliography{bibliography}

\end{document}